\documentclass[aps,pra,superscriptaddress,twocolumn,showpacs]{revtex4-2}
\usepackage[utf8]{inputenc}
\usepackage{amsmath}
\usepackage{amssymb}
\usepackage{amsfonts}
\usepackage{amstext}
\usepackage{amssymb}
\usepackage{amsthm}
\usepackage{bbold}
\usepackage{mathtools}
\usepackage{physics}
\usepackage{graphicx}
\usepackage{dcolumn}
\usepackage{bm}
\usepackage{braket}
\usepackage{color}
\usepackage{enumerate}
\usepackage[english]{babel}
\usepackage{xcolor}
\usepackage{setspace}
\usepackage[perpage, hang]{footmisc} 

\usepackage{pdfpages}
\usepackage{pgffor}

\makeatletter
\AtBeginDocument{\let\LS@rot\@undefined}
\makeatother

\DeclareUnicodeCharacter{2212}{\ensuremath{-}}

\newcommand\sucp[1]{p_{\mathrm{suc.}}}

\allowdisplaybreaks
\advance\parskip 0.4pt

\usepackage{hyperref}
\hypersetup{
	colorlinks=true,
	linkcolor=blue,
	filecolor=blue,
	citecolor = red,      
	urlcolor=cyan,
 }
\newcommand{\Autoref}[1]{%
  \begingroup%
  \def\chapterautorefname{Ch.}%
  \def\sectionautorefname{Sec.}%
  \def\subsectionautorefname{Subsec.}%
  \def\subsubsectionautorefname{Subsubsec.}%
  \def\paragraphautorefname{\P}%
  \def\tableautorefname{Table}%
  \def\equationautorefname{Eq.}%
  \def\figureautorefname{Fig.}%
  \autoref{#1}%
  \endgroup%
}

\newcommand{\figref}[1]{Fig.~\ref{#1}}
\newcommand{\appref}[1]{App.~\ref{#1}}

\newcommand{\pll}[2]{\frac{\partial #1}{\partial #2}}

\newcommand{\plm}[3]{\frac{\partial^#3 #1}{\partial #2^#3}}

\usepackage[normalem]{ulem}

\begin{document}
\setstretch{1.08}
\title{Quantum Solvers: Predictive Aeroacoustic \& Aerodynamic modeling}
	
	\author{Nis-Luca van Hülst}
   \affiliation{Institute for Quantum Physics, Department of Physics,
	University of Hamburg, Luruper Chaussee 149, 22761 Hamburg, Germany}
   \author{Theofanis Panagos}
   \affiliation{Institute for Quantum Physics, Department of Physics,
	University of Hamburg, Luruper Chaussee 149, 22761 Hamburg, Germany}
    \author{Greta Sophie Reese}
   \affiliation{Institute for Quantum Physics, Department of Physics,
	University of Hamburg, Luruper Chaussee 149, 22761 Hamburg, Germany}
   \author{Shahram Panahiyan}
   \affiliation{Institute for Quantum Physics, Department of Physics, University of Hamburg, Luruper Chaussee 149, 22761 Hamburg, Germany}
   \affiliation{Max Planck Institute for the Structure and Dynamics of Matter, Luruper Chaussee 149, 22761 Hamburg, Germany.}
    \author{Tomohiro Hashizume}
   \affiliation{Institute for Quantum Physics, Department of Physics,
	University of Hamburg, Luruper Chaussee 149, 22761 Hamburg, Germany}
   \affiliation{Max Planck Institute for the Structure and Dynamics of Matter, Luruper Chaussee 149, 22761 Hamburg, Germany.}
   \affiliation{Center of Ultrafast Imaging, Luruper Chaussee 149, 22761 Hamburg, Germany}

\date{5 November 2024}
\maketitle
\onecolumngrid
\section{Preamble}
The following pages comprise our original Phase II submission for the \emph{Airbus} x \emph{BMW Group Quantum Computing Challenge 2024}. To maintain the strict authenticity of the original competition entry, this content is presented in its exact handed-in format, commencing on the next page, aside from the specific citations to our own work, which were unpublished at the time of submission but now refer to the two following preprints:
\begin{itemize}
  \item Siegl \emph{et~al.}, “Tensor-Programmable Quantum Circuits for Solving Differential Equations” (2025) \cite{Siegl2025}
  \item van Hülst \emph{et~al.}, “Quantum-Inspired Tensor-Network Fractional-Step Method for Incompressible Flow in Curvilinear Coordinates” (2025) \cite{vanhuelst2025}
\end{itemize}
Together, these publications substantiate the technical soundness and broader impact of the methods documented in this submission.

\newpage
\twocolumngrid

\section{Submission Summary}
\begin{figure}[t]
      \centering
      \includegraphics[scale=1.00]{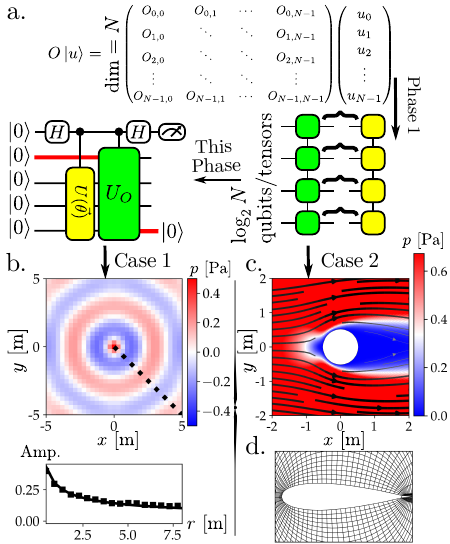}
      \caption{\label{fig:summary}
         {\bf Case I: }
         {\bf a.~Translation of matrix-vector operation to a quantum circuit.}
         In Phase 1, a field $\ket{u}$ and an operator $O$ in matrix-vector form (top) 
         is compressed into MPSs and MPOs (yellow and green, bottom right). 
         In phase 2 (bottom left), we translate of MPO to a quantum circuit $U_O$ for use in the realm of TPVQA
         that solves for variational problem parameterized by $\vec{\theta}$. 
         Here, the unitary $U(\vec{\theta})$ encodes the time evolution of $\ket{u}$ when its optimal. 
         {\bf b.~Simulation of the $2D$ wave equation with TPVQA.}
         (top) Pressure field at $t=0.1$ [s]. 
         (bottom) Local pressure amplitudes along the indicated diagonal in the top panel (square)
         and their analytical predictions (solid). 
         {\bf Case II: }
         {\bf c.~Flow around a cylinder.}
         The steady-state flow around an immersed cylinder simulated with quantum-inspired method.
         {\bf d.~Grid generation with the quantum-inspired approach.}
         Body-fitted grid of NACA0012 aerofoil, generated with our quantum-inspired method. 
         Details of the simulations in this figure are provided in the text, \appref{app:PC1numdetail}, and \appref{app:NSCC}.
      }
   \end{figure}

Recent decades have witnessed outstanding advances in technological and theoretical aspects of quantum computing
\cite{jaksch2000,bluvstein2022,bluvstein2024,evered2023}. 
In particular, quantum computing platforms are advancing at such a pace that developing quantum algorithms 
and circuits for implementing problems have become a priority \cite{bluvstein2024,googleQE2024}. 
Motivated by this, in Phase $1$ \cite{Phase1},
which is an integral part of the manuscript and is assumed to have been read,
we proposed a quantum-inspired method for solving the Navier-Stokes equations. 
We have shown that our method requires exponentially smaller computational resources $\mathcal{O}(\log_2(N))$
compared to conventional 
computational fluid dynamics (CFD) methods performed on $N$ grid points. 
In this submission, we further extend our proposed methodology in two ways: 

I) We extend our method from Phase 1 to a quantum algorithm|Tensor-Programmable Variational Quantum Algorithm (TPVQA) \cite{Siegl2024}|by 
directly translating a general Matrix Product Operator (MPO) $O$ 
\cite{schollwoeckDensitymatrixRenormalizationGroup2011a} 
into a set of unitary quantum gates $U_O$ \cite{termanovaTensorQuantumProgramming2024} 
(\Autoref{fig:summary}a.,b.)
within a Variational Quantum Algorithm (VQA) setting %
\cite{peruzzoVariationalEigenvalueSolver2014,mccleanTheoryVariationalHybrid2016,
mcardleVariationalAnsatzbasedQuantum2019,lubaschVariationalQuantumAlgorithms2020,cerezoVariationalQuantumAlgorithms2021}. 
We test its performance by solving  
problem case 1---the point source propagation---on an (emulated) quantum computer.
With this, we were able to reproduce the local pressure amplitudes which agreed perfectly with the analytical solution
(\Autoref{fig:summary}b.). 
This approach enables programming quantum algorithms using the flexible operator representation through MPOs, 
while only requiring to measure $\mathcal{O}(1+\lceil \log_2 \zeta \rceil)$ qubits, 
where $\zeta$ is the maximum bond dimension of the MPOs.
With this method, we further improve the scalability by polynomial in $\log_2 N$.

II) We extend our method from Phase $1$ 
for solving problem case $2$---a flow around an immersed object---using curvilinear coordinate method (\Autoref{fig:summary}c.,d.), allowing for an overall algorithmic complexity of $\mathcal{O}(\chi^3)$. Here $\chi$ denotes the maximum bond dimensions of the fluid fields in the Navier-Stokes equations. 
We test its performance by simulating a flow around a 
\textit{cylinder} (\figref{fig:summary}c.), demonstrating that our Matrix Product State(MPS)-based method gives accurate 
results while allowing for strong compression. 
We conducted a benchmark with industrial-grade CFD solvers and obtained a relative error of less than $0.1\%$ with a compression factor $>10$ for the horizontal force exerted on the cylinder. 
We extended our method for generating grids of general shapes. In \figref{fig:summary}d., we show a generated grid around NACA0012 aerofoil \cite{ladson1988effects}.
These methods can readily be adapted to TPVQA for a a speed-up of $\mathcal{O}(\mathrm{poly}(\chi))$. %
The integration of time evolution and grid generation is a focus of our future research. 

Our proposal offers an exponential reduction in required computational resources,
hence enables larger-scale CFD simulations.
It presents industrial opportunities such as faster optimization, design processes, and data analysis. 
Furthermore, it reduces the environmental impact of computing
by reducing CO\textsubscript{2} emissions, thus contributing to sustainability. 
Considering that with just of order $100$s of error-corrected logical qubits, 
the computational ability of a quantum computer surpasses that of the world's largest supercomputer, 
thus, further advancements in quantum computing technologies are anticipated \cite{hpcwireCrossingQuantum}. 
This can be seen in 
the experimental realizations of error-corrected logical qubits \cite{bluvstein2024,googleQE2024}, indicating that
we are at the dawn of the era of fully utilizing the exponential advantage of quantum computers.

\newpage
\cleardoublepage
\section{Detailed Explanation}

In this phase, we solve both problems from case 1 and case 2 of the challenge 
using a quantum-classical hybrid algorithm for case 1 and a quantum-inspired approach for case 2, 
by extending the proposed methodology in phase 1 \cite{Phase1} for further efficiency and accuracy. 

\textit{Problem case 1:}
In this section, we study the propagation of point source using our newly developed Tensor-Programmable VQA (TPVQA) 
\cite{Siegl2025}, which leverages the flexible operator representation through 
MPOs integrated with a powerful VQA framework. 
We show that our approach yields a highly accurate evolution of a fluid field with measurement costs that do not increase with the discretization size.

\textit{Problem case 2:}
In this section, we investigate the steady flow around an immersed cylinder. 
We use a quantum-inspired MPS-based approach that utilizes curvilinear coordinates method. 
We demonstrate that this approach provides an efficient implementation of this conventional and widely adopted CFD method.
Subsequently, we present the results of our approach for the immersed cylinder and compare their accuracy with 
those obtained from industrial CFD software.
This approach is readily translatable into a quantum algorithm through TPVQA.

\subsection{Problem case 1}\label{PC1}
\emph{Description of the problem.---}
We start by introducing the Linearized Euler Equations (LEE) for inviscid fluid in $2D$ with a point source $F(x,y,t)$ and an absorbing boundary layer $\gamma(x,y)$  (\appref{app:sponge})
\cite{israeliApproximationRadiationBoundary1981,maniReflectivitySpongeZones2010,maniReflectivitySpongeZones2010,
pericAnalyticalPredictionReflection2018,carmignianiOptimalSpongeLayer2018} as we solved in Phase $1$ \cite{Phase1}
with quantum-inspired approach
\begin{align}
    \label{eq:Euler_eqs}
    \partial_t p &= 
    - \bar{\rho}c^{2}\left(\partial_x u + \partial_y v\right) 
    - \bar{u}\partial_x p + F(x,y,t) - \gamma(x,y)p, \nonumber \\
    \partial_t u &= - \partial_x p / \bar{\rho}
    - \bar{u}\partial_x u  - \gamma(x,y)u, \nonumber\\
    \partial_t v &= - \partial_y p / \bar{\rho}
    - \bar{u}\partial_x v  - \gamma(x,y)v.
\end{align}
Here, $p=p(x,y,t)=p(t)$ [Pa] is the pressure field fluctuation 
\footnote{Here, for the sake of brevity, we use $p(t)$ instead of $p(x,y,t)$},
$u$ ($v$) [m/s] is the velocity field fluctuation component in the $x$-($y$-)directions, 
$\bar{\rho}=1.225$ [kg/m\textsuperscript{3}] is the mean density, $c=340.2$ [m/s] is the velocity of sound, and 
$\bar{u}$ is the base flow in $x$-direction. The point spurce is given by $F(x,y,t) = \frac{2c^{2}A_{0}}{2\pi f}\delta (x,y)\sin(2\pi f t)$ in which
$f=100$ [Hz] is the frequency of the source and $A_{0}=1$ [db] is the source amplitude. 

To show that the equations of this class are solvable with our quantum-classical hybrid approach, 
we focus on the limit of $\bar{u}=0$ in which the LEE equation reduces to a wave equation 
\begin{align}
   \label{eq:Euler_eqs_with_sponge}
   \partial_{t}^2p(t) = c^2 (\partial_{x}^2+\partial_{y}^2)p(t) - \gamma(x,y)\partial_{t}p(t)+
   \partial_{t}F(x,y,t). 
\end{align}
In order to solve \Autoref{eq:Euler_eqs_with_sponge} numerically, 
we first discretize the field $p$ 
in equally spaced grid points in the interval of $-L$ and $L$ [m]
with increments $\Delta x$ [m] and $\Delta y$ [m] in $x$- and $y$-directions, 
and in time with the interval $\Delta t$ [s]. 
This yields an iterative equation for evolving
$p(t)$ by time $\Delta t$ 
\begin{align}
   \label{eq:discretized}
   &p(t+\Delta t) \nonumber\\
   &=
   \left(2\mathbb{I} + (\Delta t)^2\left(c^2 O_{\partial_{x}^{2}}^{\kappa}  + c^2 O_{\partial_{y}^{2}}^{\kappa} 
   - \gamma(x,y) \right)\right) p(t) \nonumber\\
    &\ + \left( \mathbb{I} - \Delta t \gamma (x,y)  \right)p(t-\Delta t) 
    + (\Delta t)^2\partial_{t}F(x,y,t) \nonumber\\ 
    &=\mathcal{M}_1^{\kappa} p(t)
    +\mathcal{M}_2 p(t-\Delta t)
    + (\Delta t)^2\partial_{t}F(x,y,t),
\end{align}
where $O_{\partial_{x}^{2}}^{\kappa}$ and $O_{\partial_{y}^{2}}^{\kappa}$ are discretized using a central finite difference stencil of order $\kappa$
with Dirichlet boundary conditions, and $\mathbb{I}$ is an identity operator.

\begin{figure*}
   \centering
   \includegraphics[scale=1.0]{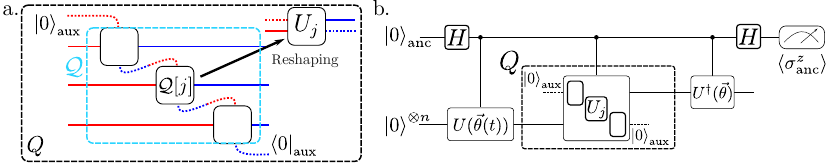}
   \caption{{\bf a.~MPO tensors to unitaries.} 
      $j$\textsuperscript{th} unitary $U_j$ of dimension $2\times 2^z$ 
   is recovered by treating the combined left bond and an incoming physical leg 
   of the $j$\textsuperscript{th} tensor of $\mathcal{Q}$, $\mathcal{Q}[j]$, as the incoming dimension of $U_j$ (red), 
   and the combined right bond and outgoing physical leg of $\mathcal{Q}[j]$ as the outgoing dimension of $U_j$ (blue).
   After projecting out the auxiliary dimension in $Q$ (black box), $\mathcal{Q}$ is recovered (light blue). 
   {\bf b.~VQA with phase kickback algorithm.}
   We evaluate the expectation values in the cost function by performing the Hadamard test with the unitary 
   $U(\vec{\theta}(t))QU^\dag(\vec{\theta})$. 
   The projection of auxiliary the qubit is done by post-selecting the runs with successful measurement 
   of $\ket{0}_{\mathrm{aux}}$. 
   \label{fig:unitarization}}
\end{figure*}

\emph{Tensor-Programmable Variational Quantum Algorithm.---}
In order to evolve the field $p(t)$ utilizing a  quantum computer,
we propose a method that integrates the MPO-quantum circuit translation technique \cite{termanovaTensorQuantumProgramming2024} 
to the framework of VQA
\cite{peruzzoVariationalEigenvalueSolver2014,mccleanTheoryVariationalHybrid2016,
mcardleVariationalAnsatzbasedQuantum2019,lubaschVariationalQuantumAlgorithms2020,cerezoVariationalQuantumAlgorithms2021},
 i.e.~TPVQA. VQA is a classical-quantum hybrid algorithm that solves a variational problem by utilizing 
$n$-qubit quantum state that is encoded by an ansatz circuit $U(\vec{\theta})$ that consists 
of multi-qubit gates and single qubit rotation gates parameterized by $\vec{\theta}$ (\figref{fig:unitarization}b.).
While it classically solves a variational problem, a cost function is evaluated with a quantum computer. 

For our simulation, we formulate the evolution of $p$ as a variational problem.
First let $\ket{p(t)}$ be the encoding of the field values of $p$ onto the amplitude of $2^{n}$ 
basis states in Hilbert space spanned by $n=\log_2 (4L^2/(\Delta x \Delta y))$ qubits, up to the global normalization factor $\lambda_p(t)$
(\appref{app:ampenc}). 
Based on \Autoref{eq:discretized}, given that we know the optimal sets of parameters for encoding 
$\ket{p(t)}=\lambda_p(t)U(\vec{\theta}(t))\ket{0}^{\otimes n}$ and 
$\ket{p(t-\Delta t)} = \lambda_p(t-\Delta t)U(\vec{\theta}(t-\Delta t))\ket{0}^{\otimes n}$ 
to a quantum state, a parameter set for $U(\vec{\theta})$ that encode $\ket{p(t+\Delta t)}$ in the form 
$\ket{p(t+\Delta t)}=\theta_{0}U(\vec{\theta}(t+\Delta t))\ket{0}^{\otimes n}$ 
to a quantum state is obtained by solving the minimization problem with the following cost function
\begin{align}
   \mathcal{C}(\theta_0,\vec{\theta}) =& 
   \Big|\Big|\theta_0 U(\vec{\theta})\ket{0} -\lambda_p(t)
   \mathcal{M}_1^{\kappa}
   \ket{p(t)} \nonumber\\
                                       &- \lambda_p(t-\Delta t) \mathcal{M}_2 \ket{p(t-\Delta t)} \nonumber\\
               &- 2c^2A_0\cos(2\pi f t) \sigma^x_{n_x}\sigma^x_{n_y}\ket{0}\Big|\Big|^2,
\end{align}
for parameters $\theta_0$ and $\vec{\theta}=\{\theta_1,\theta_2, \ldots\}$.

In VQA, the terms in the cost function of the form, 
$\Re \{ \bra{0}U^\dag(\vec{\theta})\mathcal{M} U(\vec{\theta}(t))\ket{0} \}$
are evaluated using a quantum computer. 
However, our operators $\mathcal{M}^\kappa_1$ and $\mathcal{M}_2$ are not, in general, completely positive trace-preserving maps (CPTPMs).
Therefore, we must first find representations of them embedded in CPTPMs in a larger Hilbert space, then project out unnecessary dimensions.

Here, we find such a representation by finding a unitary that closely approximates a general 
matrix $\mathcal{M}$ (e.g.~$\mathcal{M}^{\kappa}_1$ and $\mathcal{M}_2$)
via quantum-tensor programming \cite{termanovaTensorQuantumProgramming2024}, the TP part of our TPVQA. In Phase 1 \cite{Phase1}, we showed how a matrix $\mathcal{M}\in \mathbb{C}^{(2^{n}\times2^{n})}$ 
can be exponentially compressed into an MPO, 
a length $n$ chain of tensors with dimensions of at worst $\zeta \times 2 \times 2\times \zeta$, 
by extending amplitude encoding to general matrices, where $\zeta$ is the maximum bond dimension (\appref{app:MPS-MPO}). 
Here, we propose a method for translating  and integrating $\mathcal{M}$ in MPO representation into a realm of VQA 
by finding $n$ unitary operators of dimension $2\times2^{z}>2\zeta$ 
of the form $U_j = I_{0}\otimes I_{1} \cdots \otimes I_{j-1} \otimes U_{j} \otimes I_{j+1} \otimes \cdots I_{n+z}$
that act on $n+z$ qubits. The $I_i$ is an identity acting on a local Hilbert space,
such that $Q=\prod_j U_j$ approximates $\mathcal{M}$
up to a normalization factor $a_q$ and a projection of $z$ auxiliary qubits. 

$Q$ is obtained by first finding a unitary MPO $\mathcal{Q}$ 
with every bond being $2^{z}$ such that it approximates $\mathcal{M}$. 
This is done by solving a minimization problem with the following cost function 
\begin{align}
   C(a_q,\mathcal{Q})=%
    \  \lVert a_q \mathcal{Q} -\mathcal{M} \rVert^2, \label{eq:minimization_mpo}
\end{align}
on a classical computer, under a condition that reshaped tensors of $\mathcal{Q}$ reside in Stiefel manifold (\appref{app:MPOQC}). 
Here, the tensors in $\mathcal{Q}$ and unitaries in $Q$ are related via a simple reshaping and projection 
$\mathcal{Q} = \ket{0}_{\mathrm{aux}}\bra{0}_{\mathrm{aux}} Q \ket{0}_{\mathrm{aux}}\bra{0}_{\mathrm{aux}} $ 
\Autoref{fig:unitarization}a.
For the operators $\mathcal{M}_1^{\kappa}$, $\mathcal{M}_2$, and similar differential operators, 
the MPO bond dimensions required to express them only depend on $\kappa$, and they do not scale with $n$ (\appref{app:MPS-MPO}).
For translating $\mathcal{M}_1^{\kappa}$ and $\mathcal{M}_2$, only $z=4$ auxiliary qubits were required.

Once the translation is done, we evaluate the expectation value $\Re\{ \bra{0}U^\dag(\vec{\theta})\mathcal{M} U(\vec{\theta}(t))\ket{0} \}=a_q\Re\{\bra{0}U^\dag(\vec{\theta})P_{\ket{0}_{\mathrm{aux}}\bra{0}_{\mathrm{aux}}}QU(\vec{\theta}(t))\ket{0}\}$ using the quantum network depicted in Fig. \ref{fig:unitarization}b. %
Here, we use a Hadamard test \cite{aharonovPolynomialQuantumAlgorithm2009,lubaschVariationalQuantumAlgorithms2020}, which transfers the latter expectation value into the $\braket{\sigma^{z}}_{\mathrm{anc}}$ of the ancilla, upto a prefactor. Since we project out the auxiliary qubits in $\ket{0}_{\text{aux}}$, defining a successful application of matrix $\mathcal{M}$ with probability $P_{\text{succ}}$, this prefactor becomes dependent on $P_{\text{succ}}$ (\appref{app:hadamardtest}).
Thus, an evaluation of the cost function only requires measurements on three circuits, 
offering exponentially fast evaluation of vector-matrix-vector products for the cost function evaluation. 

\emph{Results: simulation with TPVQA.---}
Now we show the applicability of our approach by evolving the pressure field $p$
on a classically emulated noise-free quantum computer. 
We assume that the Ansatz circuit contains the solution, 
the measured expectation values are exact, and the classical optimizer provides the global optima. 
In \Autoref{fig:summary}b.~we show the field  at $t=0.1$ [s] performed with $\kappa=6$\textsuperscript{th}-order stencil, 
where the bottom panel shows the perfect agreement between the numerically obtained local amplitudes of the oscillations
and those of the analytically derived values (\appref{app:pc1analytical}). 
We also provide an animation of the time evolution up to $t=0.1$ [s] as supplementary material. 

\begin{figure}
   \includegraphics[width=\columnwidth]{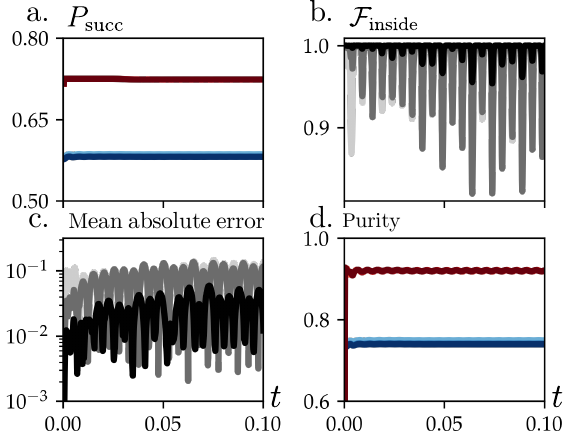}
   \caption{
   \label{fig:sprobandpurityevol}
   {\bf a.~ Evolution of success probabilities ($P_{\mathrm{succ}}$).} 
   Shown are $P_{\mathrm{succ}}$ for evaluating $\Re\{\bra{p(t+\Delta t)}Q_1\ket{p(t)}\}$ (blue) 
   and $\Re\{ \bra{p(t+\Delta t)}Q_2\ket{p(t-\Delta t)}\}$ (red). 
      {\bf b.~Fidelities.} 
      Fidelities between the evolved state and numerically exact result
      excluding the boundary sponge layer after imposing the normalization $\mathcal{F}_{\mathrm{inside}}$.
      {\bf c.~Mean absolute error.} 
      Mean absolute error of an evolved state to the numerically exact result inside the boundary sponge layer. 
      {\bf d.~Evolution of the purity of the ancilla.}
      Shown are the purities of the ancilla for computing $\bra{p(t+\Delta t)}Q_1\ket{p(t)}$ (blue) and 
      $\bra{p(t+\Delta t)}Q_2\ket{p(t-\Delta t)}$ (red). 
      All quantities are computed for simulations using 
      $\kappa=2$\textsuperscript{nd},$4$\textsuperscript{th}, and $6$\textsuperscript{th} order 
      central finite difference stencil 
      for discretizing the differential operators (light to dark). 
   }
\end{figure}

We then, in \Autoref{fig:sprobandpurityevol}a.,
analyze the evolution of $P_{\mathrm{succ}}$ for evaluating the expectation values with 
$Q^\kappa_1$ (blue) and $Q_2$ (red), which are the unitaries derived from MPOs 
$\mathcal{Q}^\kappa_1$ and $\mathcal{Q}_2$. 
We see that they immediately saturate to a $\kappa$-independent value, and they do not vanish. 
Shown in \Autoref{fig:sprobandpurityevol}b., 
are fidelities $\mathcal{F}_{\mathrm{inside}}$ 
between the region excluding the sponge layer of the simulated state 
and the numerically exact state obtained with exact matrix-vector simulation, after the normalization. 
In \Autoref{fig:sprobandpurityevol}c., we also provide their mean absolute errors. 
Here we see that the fidelity gets better as the order of the stencil becomes higher, 
compensating the error introduced by the translation stage of TPVQA. 
Use of $\kappa$-accurate stencil only costs at worst $\mathcal{O}(\kappa)$ additional
MPO bond dimensions and at worst $\mathcal{O}(\log_2 \kappa)$ additional auxiliary qubits, 
thus, exponentially accurate evaluations of derivatives are possible using TPVQA, 
with a mitigation of translation error that comes for free.

In a realistic case, one must evaluate the cost function by evaluating the expectation values 
via the measurement of the ancilla. 
The efficiency of this measurement depends on the purity of the ancilla, and this is shown in \Autoref{fig:sprobandpurityevol}c.
For small $\Delta t=10^{-4}$ [s], $\mathcal{M}^{\kappa}_1$ and $\mathcal{M}_2$ are dominated by the identity, 
and hence the phase kickback does not introduce large entanglement between the state and the ancilla. 
As a result, the purity of the ancilla at the minima of $\mathcal{C}(\theta_0,\vec{\theta})$ stays high throughout the evolution (see \Autoref{fig:sprobandpurityevol}d). 

Our approach, for fixed resolution of the expectation values, 
scales like $\mathcal{O}(\log_2(\kappa))$ for the number of measurements, 
scales like $\mathcal{O}(n+\log_2 \kappa)$ for the number of qubits,
and at worst $\mathcal{O}(n\kappa^2)$ scaling for the circuit depth \cite{nielsen_chuang2010}.
Furthermore, for data structures that only require area-law entangled (MPS-like) quantum states 
for their amplitude encoding, it is known that the required circuit depth, and hence the number of variational parameters
of $U(\vec{\theta})$, is $\mathcal{O}(\mathrm{poly}(n))$ 
\cite{lubaschVariationalQuantumAlgorithms2020}. 
Thus, TPVQA provides us with further speedup and scalability to our proposed MPS-MPO based methods in Phase 1. 

\emph{Conclusion.---}
In this section, we have introduced TPVQA for translating MPOs into quantum circuits and embedded it in the powerful framework of VQA. By directly translating the time evolution operator---represented as MPO---into a quantum circuit, 
we showed that our method provides substantial improvements in terms of resources for the cost function evaluation and stability of time evolution compared to conventional term-by-term evaluation \cite{lubaschVariationalQuantumAlgorithms2020}.
Overall, our approach provides a flexible and easy-to-use framework for solving a wide range of PDEs with a quantum computer, including nonlinear PDEs, systems with higher dimensions, and systems with complicated geometries, including the case of solving the Navier-Stokes equations on body-fitted coordinates, which its MPS-MPO fomrulation is discussed in the next section.
The details on the numerical simulations are given in \appref{app:PC1numdetail}.

\subsection{Problem case 2}\label{PC2}
\emph{Description of the problem.---} 
In this section, we solve for the flow around an immersed cylinder using the curvilinear coordinate method on a body-fitted coordinate system.
The concept can be readily applied to other cases if a suitable grid is used and the solver is switched to a compressible solver. 
We note that this switch will have no influence on the scaling with respect to the bond dimension of the flow fields.

The $2D$ incompressible Navier-Stokes equations (ICNS) are given by \cite{DEFRUTOS2016}
\begin{align}
    \nabla \cdot \bm{u} &= 0, \\
    \partial_t \bm{u}+(\bm{u} \cdot \nabla)\bm{u} &= -\frac{\nabla p}{\rho} + \nu \Delta \bm{u}, \label{eq:mom-eq}
\end{align}
where $\bm{u}=(u, v)$ is the velocity field with components $u$, and $v$ in $x$ and $y$-directions, $p$ is the pressure,
$\rho=1$ [kg/m\textsuperscript{3}] is the density, and $\nu$ [Pa$\cdot$s] is the kinematic viscosity.
The flow fields are defined on an orthogonal computational grid, $\xi(x,y)$ and $\eta(x,y)$,
where each point $(\xi,\eta)$ corresponds bijectively to a point in the physical space $(x,y)$. 
Furthermore, we define $\eta$ such that the $\eta=0$ coordinate line maps to the surface of the immersed body, with $\eta=1$ corresponding to the far field boundary, and $\xi$ that is linearly independent with $\eta$ with periodic boundary condition along a branch cut  (\Autoref{fig:compgrid} in \appref{app:NSCC}).

The fields $u(\xi, \eta)$, $v(\xi, \eta)$, and $p(\xi, \eta)$ 
are encoded in the amplitudes of a state vector as described in \appref{app:ampenc}. 
This state vector is then represented as a MPS with a maximum bond dimension of $\chi$.
The differential operators in computational grid are discretized using finite differences and expressed as MPOs (\appref{app:MPS-MPO}), which allows an exponential compression in their presentations as we discussed in Phase 1 \cite{Phase1}. 
\emph{Methods.---}\label{sec:Methods}
To solve the ICNS using curvilinear coordinates \cite{ZANG199418, Ferziger2002:CMFD}, 
we first find a transformation of derivative operators, changing from derivatives with respect to $x$ and $y$ to derivatives with respect to $\xi$ and $\eta$. For instance, the partial derivative $\partial_{x}$ is expressed as \cite{Anderson1995}
\begin{align}
    \partial_{x} = \frac{1}{J}\bigg[ \underbrace{(\partial_{\eta} y)}_{\textit{metric}} \partial_\xi - \underbrace{(\partial_\xi y)}_{\textit{metric}} \partial_\eta \bigg] \label{eq:del_x_curvi},
\end{align}
where $J:=J(\xi,\eta)$ denotes the determinant of the Jacobian of the coordinate transform and $\partial_{\xi}$ ($\partial_{\eta}$) denote partial derivatives with respect to $\xi$ ($\eta$), respectively. Similar curvilinear expressions can be found for $\partial_{y}$ and $\Delta$ \cite{Anderson1995}. %

We discretize the physical domain by introducing a set of grid coordinates $(\bm{x}, \bm{y}):=\{(x_i, y_i)|\ i=0,..,2^n-1\}$, where $n=n_\xi + n_\eta$ and $n_\xi$ ($n_\eta$) denotes the number of tensors for the $\xi$-($\eta$-)direction. We amplitude-encode (\appref{app:ampenc}) the grid coordinates $\bm{x}$ and $\bm{y}$ into an MPS $\ket{\bm{x}}$ and $\ket{\bm{y}}$, respectively. The \textit{metrics} (see \Autoref{eq:del_x_curvi}) are then computed by applying Cartesian finite difference operators $\partial_\xi$ ($\partial_\eta$) to $\ket{\bm{x}}$ and $\ket{\bm{y}}$ via crude MPO-MPS contractions. Returning to the example of $\partial_{x}$, we require to combine expressions as $(\partial_{\eta} y) \partial_\xi$, where \textit{metric} $\partial_{\eta} y$ is represented as an MPS and $\partial_\xi$ is an MPO, to form a new MPO. This operation scales as $\mathcal{O}(\chi_{met}^2)$ (\appref{app:Scaling}), where $\chi_{met}$ is the maximum bond dimension of an MPS representing one of the \textit{metrics}.

With the curvilinear differential operators encoded as MPOs, incorporating geometric information, and the fluid fields represented as MPSs, we evolve the system in time by using Chorin's projection method \cite{Chorin1968, DEFRUTOS2016} (\appref{app:Chorins})
as an explicit time integrator.
In this method, the velocity field  $\bm{u}^{(k)}$ at time step $k$ is first advanced to an intermediate state $\bm{u}^*$ using the momentum equations (\Autoref{eq:mom-eq}), neglecting the pressure gradient. This step scales with $\mathcal{O}(\chi^3)$ \cite{Michailidis2024} (\appref{app:Scaling}). Next, we must solve a Poisson equation to determine the pressure $p^{(k+1)}$ at the subsequent time step $k+1$. In this equation, the source term is proportional to $\nabla \bm{u}^*$. %
This step also scales with $\mathcal{O}(\chi^3)$ \cite{Oseledets2012} \appref{app:Scaling}).
Lastly, the velocity is projected back onto the divergence-free manifold. This is done via a plain MPO-MPS contraction and a MPS-MPS addition, which is negligible compared to former steps \cite{schollwoeckDensitymatrixRenormalizationGroup2011a,paeckelTimeevolutionMethodsMatrixproduct2019}. Overall, this results in a total complexity for the algorithm of $\mathcal{O}(\chi^3)$.

\begin{figure}
   \centering
   \includegraphics[width=1.0\columnwidth]{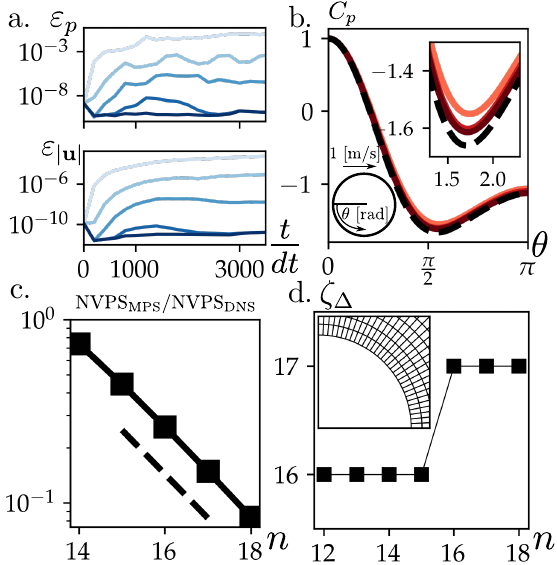} %
   \caption{
      \textbf{Simulation of a flow around a cylinder:}
      \textbf{a.~Relative errors of the simulations.} 
      $\epsilon_{p}$ (top) and $\varepsilon_{|\mathbf{u}|}$ (bottom)
   between MPS-based and DNS simulations as a function of the time steps for different 
   bond dimensions of $20$, $30$, $40$, $50$, and $60$ (light to dark), respectively, 
   for a fixed system size of $n_\xi=n_\eta=7$. 
   \textbf{b.~Validation with a reference solution.} 
   Shown are $C_p$ of the steady state along the cylinder for $n\in\{14, 16, 18\}$ (light to dark)
   qubits per spatial dimension with $\chi=35$. 
   This is compared to a reference solution computed with FVM with $2^{20}$ cells.
   In the inset, we enlarge near $\theta=\pi/2$ where the convergence towards the reference solution is most apparent.
   \textbf{c.~The compression rate of our approach.} 
   The fraction of the number of variables parametrizing solutions for MPS-based and DNS solutions as a function of $n$. We obtain the scaling of the compression rate that scales like $\propto (2^n)^{-0.8}$ (dashed)
   Simulations are done for an immersed cylinder with a radius $r=0.5$ [m] 
   in a fluid with $\nu=0.05$ and $U_\infty=1$[m/s].
   \textbf{Grid generation: }
   \textbf{d.~The size of curvilinear Laplacians.} Maximum bond dimension $\zeta_{\Delta}$ for the curvilinear Laplacian MPO with Dirichlet boundaries on our cylinder grid, up to discarding singular values less than $10^{-15}$ at each bond.
   }
   \label{fig:l2errAndCpchi=35}

\end{figure}

\emph{Results: flow around immersed cylinder---}
We present the results of our MPS-based simulations using the example of laminar flow around a cylinder. We consider a cylinder with radius $r=0.5$ [m], which is immersed in a fluid characterized by a viscosity of $\nu=0.05$ [Pa$\cdot$s] and a uniform stream velocity $U_\infty=1$ [m/s] in horizontal ($x_1$) direction \cite{Ferziger2002:CMFD}. The physical domain is discretized using a transfinite interpolated \cite{Gordon1982} collocated O-grid---extending outwards to a distance of $r=8$ [m] in the far field---generated via MPS from the surface and boundary point distribution. 
The grid and its relation to the computational domain are depicted in \Autoref{fig:compgrid} in \appref{app:NSCC}.
The curvilinear differential operators for this grid are constructed as explained in the previous \textit{Methods} section, exclusively in the MPS picture.
We employ finite difference stencils with a second-order accuracy unless specified otherwise.

In the first study, we investigate the effects of the bond dimension on the accuracy of the MPS-based simulations. Fixing our system size to $n_\xi=n_\eta=7$, translating to a grid size of $2^7 \times 2^7$, we aim to quantify the
accuracy of our MPS-based simulation on the transient path towards the steady-state, compared to a Direct Numerical Simulation (DNS). For this purpose, we introduce the relative error  $\varepsilon_{\mu}=\frac{||\mu - \mu_{DNS}||_2}{||\mu_{DNS}||_2}$, where $\mu\in\{p,|\mathbf{u}|\}$. 
The results are presented in \Autoref{fig:l2errAndCpchi=35}a. The accuracy saturates as a function of the time steps irrespective of choices for the bond dimension and the quantity under investigation. From a small bond dimension ($\chi=20$) to a larger bond dimension ($\chi=50$), we observe an accuracy improvement of $8$ orders of magnitude. 
Further increment in the bond dimension does not affect the accuracy significantly indicating saturation in obtainable accuracy as a function of the bond dimension.  

In the second study, we aim to investigate the steady-state results in the limit of increasing system sizes quantitatively and to check whether our MPS-based solver can accurately predict steady-states despite being highly compressive. In particular, we are interested in finding out whether the intermediate dynamics leading to the steady state must be captured with high accuracy, or if its disregard through the MPS truncations is in order and predicts a steady state with correct physical quantities of practical importance. 
The \Autoref{fig:l2errAndCpchi=35}b shows the results of the MPS-based simulations (red lines)
and compares them with a finite volume method (FVM) reference (dashed). 
The results from FVM have been generated in close cooperation with our partners at the Hamburg University of Technology - Institute for Fluid Dynamics and Ship Theory with an in-house version of openFOAM \cite{Weller1998}.
For the computation, the same physical domain is discretized on a $2^{10} \times 2^{10}$ block-structured mesh.
The SIMPLE algorithm \cite{Ferziger2002:CMFD} is employed for the mass and momentum conservation when computing the laminar steady state for the above reference example. 

We compare the pressure coefficient 
$C_p$ obtained from the surface pressure distribution by evaluating the expression \cite{Clancy1986}
\begin{equation}
    C_p=\frac{p-p_{\infty}}{\frac{1}{2} \rho U_\infty^2},
\end{equation}
where $p_{\infty}$ denotes the free-stream pressure, taken from the far-field at the end of a simulation. We show the resulting $C_p$ curves in \Autoref{fig:l2errAndCpchi=35}b. 
Based on these curves, the horizontal force $F_{x,h}$ exerted on the cylinder is estimated upon numerical integration.
For $n=18$ and $\chi=35$ (green stars \Autoref{fig:l2errAndCpchi=35}b), we obtain $F_{x,h}\approx0.7097$ [N], which represents a relative error of less than $0.1\%$ compared to the reference, despite having a strong compression.
To quantify the achieved compression, we introduce the number of variables parameterizing the solution (NVPS) \cite{kiffnerTensorNetworkReduced2023} and compare its value for the MPS with that of DNS. For the MPS, the NVPS is calculated by summing the sizes of its individual tensors, of order $\mathcal{O}(2n\chi^2)$, while for the DNS it is given by $\text{NVPS}_{\text{DNS}}=2^{n}$.
The results are presented in Fig. \ref{fig:l2errAndCpchi=35}c.
Notably, for $n=18$, which represents the finest simulation in our MPS-based approach, we require only 8.4\% of the NVPS compared to DNS. This corresponds to a compression factor greater than 10. The plot further shows that the compression rate scales with $\propto (2^n)^{-0.8}$, which means exponential compression.
\emph{Results: Grid generation with MPSs and MPOs---} To investigate the flow around an immersed NACA0012 aerofoil, the grid choice is paramount for optimizing the computational effort, as it determines the maximum bond dimension $\zeta$ of the curvilinear differential operators, such as Laplacian ($\zeta_{\Delta}$). 
The choice of grid must balance several factors: first, allocating computational resources where needed \cite{ZANG199418}, 
and second, specific to MPS, ensuring that the bond dimension does not significantly increase with system size. 
There are more requirements generally, which we do not detail here \cite{vanhuelst2025}. 

A simplistic possibility is again the use of a transfinite interpolated grid as for the cylinder. However, this approach is insufficient with respect to the first mentioned aspect, especially for higher Mach numbers and more complex flow scenarios, since the boundary layers have to be resolved accurately \cite{ZANG199418}.
To overcome this drawback, various types of advanced grid generation methods have been established in the field of CFD. A well-known example is elliptic mesh generation methods \cite{Sorenson1980ACP, Thompson1982}, which solve Poisson equations to determine the grid coordinates $(\bm{x}, \bm{y})$.
Based on the algorithmic scalings shown in \appref{app:Chorins}, we conclude that the Poisson equation can be solved efficiently with a complexity of $O(\chi^3)$, where $\chi$ corresponds to the bond dimension of the MPS encoding the grid points.
We show an elliptic mesh generated using our MPS toolbox with $n_\xi=n_\eta=7$ (\Autoref{fig:summary}d). The computations on these grids are beyond the scope of this challenge and are currently part of ongoing research with our partners at Hamburg University of Technology \cite{vanhuelst2025}.

\emph{Conclusion---} 
In this section, we have underscored that the framework of MPS is general and can efficiently simulate the Navier-Stokes equations in body-fitted coordinates, adopting the widely used CFD approach with curvilinear coordinates. 
We have shown that we can accurately predict figures of merit such as the horizontal force on the cylinder while exhibiting a strong compression. Here, we have benchmarked against an industrial-grade reference. 
Furthermore, we have shown that the bond dimensions of the curvilinear Laplacian operator in the case of the cylinder stay nearly constant with increasing system size. 
\cleardoublepage
\newpage

\section{Conclusion}   \label{Con}
In Phase 1, we used a quantum-inspired approach based on MPS to simulate the noise propagation of a point source with non-reflective boundary conditions. 
We have shown that our method can be directly applied to staggered grid approaches and that all involved operators admit a fixed, low-rank representation, independent of the system size, allowing for exponential compressions for their representations. We have also seen that a crossover between the classical and quantum-inspired methods can be expected in terms of computing time for growing systems. Building on these results, in the second phase, we %
undertook to tackle two problems namely I) translating the MPS-MPO based algorithm to its corresponding quantum circuit
and II) extending the MPS approach to curvilinear coordinates on the other. The former is particularly important as it greatly simplifies access to efficient quantum algorithms and removes the burden of manually designing quantum circuits. The second underpins the versatility of the MPS-MPO framework for efficiently addressing CFD problems.

For the first problem, we have shown how to use the flexible MPO representation operators, generally for unitary or non-unitary operators, to arrive at a set of unitaries that can then be applied 
in a realistic quantum computing setup.
We have embedded these unitaries in the variational framework allowing us to overcome the bottlenecks of direct quantum algorithms, where the encoded unitaries are often applied in a deep circuit suffering from the exponentially vanishing success probabilities. 
Our method has two advantages; since we 
have the classical parameterization of the state through the variational approach, we can reproduce it at any time and determine observables of interest. 
On the other hand, unsuccessful quantum evaluations are used to estimate the success probability, which in turn is used as a norm correction in the cost term. Therefore, no resources are wasted. Overall, our MPS toolbox is a black box that can be used to transform classical matrices and vectors found in different branches of CFD into MPOs and MPSs, and subsequently to unitaries that can be utilized with VQAs. Therefore, via our method, Tensor-Programmable Variational Quantum Algorithm (TPVQA), we open up access to efficient quantum algorithms for CFD problems to a broad audience.

In the second problem, we have extended the MPS-MPO framework for solving Navier-stokes flows with immersed objects. Here, we have used the widely adopted approach of curvilinear coordinates, proving the generality and power of our MPS-MPO algorithm. We compared the simulation results with our own DNS results for a small system and the quantitative comparison in the limit of the growing system with industrial grade reference generated via openFOAM %
. We achieved remarkable accuracies, despite strong compression of the fields. We have also shown for our cylinder grid that the curvilinear operators involved are slightly larger than in the Cartesian case, but are nearly constant over the different system sizes. This corresponds to an exponential compression of the operator representation. Overall, by solving the second problem, we have underscored the capability of our MPO-MPS algorithms to undertake CFD problems with different degrees of difficulty and solve them efficiently and with substantially reduced computational cost. 

In the next step, we will further explore our method by
simulating the NACA0012 aerofoil and other complex geometries. In particular, we will focus on the grid generation part as a stand-alone tool and investigate its influence on the dimensions of the resulting generalized operators to further optimize the complexity of the algorithm. We aim to translate our MPS-MPO algorithms to quantum circuits in anticipation of the upcoming fault-tolerant error-corrected quantum computers and quantum computation platforms.

\section*{Acknowledgements}
We would like to thank Pia~Siegl (DLR) for contributing to the simulation code. Further, we would like to thank Pia~Siegl
and our supervisor Prof.~Dieter Jaksch (UHH) for fruitful discussions, and valuable comments on the implementation and interpretation of the results. We would also like to thank Paul Over and Dr.~Sergio Bengoechea from the Technical University of Hamburg-Harburg (TUHH) for assistance on the implementation of the (curvilinear) solver, provision of the reference solution and discussion on simulation results. %

\widetext
\cleardoublepage

\cleardoublepage

\bibliography{bibliography.bib}

\newpage
\appendix

\section{Sponge operators for absorbing boundary conditions}
\label{app:sponge}
We define the sponge operator $\gamma(x,y)=\gamma(x)+\gamma(y)$ 
for implementing absorbing boundary conditions in a two-dimensional Cartesian grid as 
\begin{align}
    \gamma_x(x) &= \gamma_{l, \text{max}} \ \frac{e^{(\frac{x_l-x}{x_l-x_{lb}})^{c_b}}-1}{e - 1}H(x_l-x) \notag \\
    &+ \gamma_{r, \text{max}} \ \frac{e^{(\frac{x-x_r}{x_{rb} - x_r})^{c_b}}-1}{e - 1}H(x-x_r).
\end{align}
Here $c_b>0$ is a blending coefficient, 
$\gamma_{l, \text{max}}$ ($\gamma_{r, \text{max}}$) 
is the maximum absorption rate at the left (right) boundary,
$x_l$  ($x_r$) is the $x$-coordinate of the beginning of the sponge layer towards the left (right) boundary, 
and $x_{lb}$ ($x_{rb}$) is the $x$-coordinate of the left (right) boundary. 
Although there are other implementations of non-reflective boundary conditions, we chose to use this implementation because
$\gamma(x,y)$ has a low-rank MPO representation \cite{Phase1}.

\section{Amplitude encoding} 
\label{app:ampenc} 
To propagate the discretized field $p(x,y,t)$ in time, we first discuss how it is encoded 
as a normalized quantum state $\ket{\mathfrak{p}(x,y,t)}$ on quantum hardware via 
amplitude encoding.
Amplitude encoding encodes the field values on $2^{n_x}=2L/\Delta x$ by $2^{n_y}=2L/\Delta y$ grid 
in the amplitude of the $2^{n=n_x+n_y}$ basis states spanned by $n$ qubits. 
Here, we use the computational basis states spanned by the Kronecker product of $+1$ and $-1$ 
eigenstate of Pauli-z operator $\sigma^{z}$, $\ket{0}$ and $\ket{1}$. 
Let $(i_x,i_y)$ be a point that is $i_x$\textsuperscript{th} point from the left bound, 
and $i_y$\textsuperscript{th} point from the top bound.
Then the basis state corresponding to this point is $\ket{(i_y,i_x)_b}$, where $(i_y,i_x)_b$ denotes the joint 
binary form of $i_x$ and $i_y$.
For example the field value at point $(1,5)$ for $n_x=4$ and $n_y=4$ is encoded in basis state $\ket{01010001}$. 

With this formulation, the field values of $p(x,y,t)$ are encoded to an unnormalized state $\ket{p(x,y,t)}$ as 
\begin{align}
   \ket{p(x,y,t)}=\sum_{i_x,i_y} p(x_0+i_x\Delta x,y_0+i_y\Delta_y,t)\ket{(i_y,i_x)_b},
\end{align}
where $x_0$ and $y_0$ are coordinate values of left and top boundaries.
If we tensorize the numerical representation of this state in a classical computer, 
we recover the MPS formalized of the quantum-inspired approach discussed in our first submission \cite{Phase1}.
However, we would like to encode this state into a quantum computer. 
Therefore, we find a representation of $\ket{p(x,y,t)}=\lambda_p(t)\ket{\mathfrak{p}(x,y)}$
such that $\braket{\mathfrak{p}(x,y,t)|\mathfrak{p}(x,y,t)}=1$. 
The initial states that are required for solving $\Autoref{eq:Euler_eqs_with_sponge}$ for the problem case $1$ 
can be trivially determined. 
Since we require $p(x,y,0)=0$ and $p(x,y,\Delta t) = \pll{}{t}f(x,y,t)=2c^2A_0\cos(\omega t) \ket{(2^{{n_y}-1},2^{{n_x}-1})_b}$,
we obtain $\lambda_p(0) = 0$, 
$\ket{\mathfrak{p}(x,y,\Delta t}=\sigma^{x}_{n_x+n_y-1}\sigma^{x}_{n_x-1}\ket{0}$, and $\lambda_p(\Delta t) = 2c^2A_0\cos(\omega t) $,
where $\sigma^{x}_i$ is a bit flip operator of $i$\textsuperscript{th} qubit.

\section{MPS and MPO} \label{app:MPS-MPO}
Given we are provided an (unnormalized) state $\ket{p}$ of a scalar field $p(x,y)$, for example through amplitude encoding. An MPS representation of this state $\ket{p}$ can be found by a successive SVD procedure \cite{schollwoeckDensitymatrixRenormalizationGroup2011a}. In the end, we arrive a chain of tensors $A^{i_{d}^{(j)}}:=A_{a_{j-1}, a_j}^{i_{d}^{(j)}}$ with $d \in \{ x,y\}$
\begin{align}
   \label{eq:QTTstate}
      \ket{p} = 
      \sum_{i_{x}^{(0)} ..i_x^{(n-1)}}
      \sum_{i_{y}^{(0)} ..i_y^{(n-1)}} 
      A^{i_{x}^{(0)}}..A^{i_{x}^{(n-1)}}
      A^{i_{y}^{(0)}}..A^{i_{y}^{(n-1)}}
      \ket{(i_y,i_x)_b},
\end{align}
where $i_x^{(j)}$, $i_y^{(k)} \in \{0, 1\}$  are
the $j$\textsuperscript{th} ($k$\textsuperscript{th}) digit of the binary representation of $i_x$ ($i_y$), 
and here we omitted an explicit summation over the lower indices for brevity in notation. In analogy to the qubit picture, each tensor represents one scale in our grid \cite{gourianovQuantuminspiredApproachExploit2022}. 

The creation of the MPS via successive SVD has the following steps \cite{schollwoeckDensitymatrixRenormalizationGroup2011a}: First, we reshape the vector such that the degrees of freedom $d$ are part of the number of rows. Here, the degrees of freedom equalize the number of values that the binary indices $i_x^{(j)}$ and $i_y^{(k)}$ can take, i.e.~$d=2$. Second, we find the SVD of the reshaped matrix. This results in three matrices; the isometric matrix $U$, the diagonal matrix $S$, and the isometric matrix $V^\dagger$. The matrix $S$ contains the singular values with decaying values. Third, we truncate the singular values. By cutting out the low-value singular values, we can significantly decrease the size of $U$, $S$, and $V^\dagger$. The maximal number of singular values we keep is the bond dimension $\chi$ which decides the scaling of our algorithms. Fourth, we reshape $U$ in a 3-leg tensor and add it to the MPS. We further contract $S$ and $V^\dagger$, and use them to reinitialize the procedure.

Depending on the structure of the initial data, strong truncation to low $\chi$ without or with only little loss of information is possible.
If no truncation takes place, the maximal bond dimension is $\chi_{max}=2^{\lfloor Dn/2\rfloor}$.
Many basic functions \cite{oseledets2010} are known to have efficient MPS representations.
For example, representing a constant function, 
as required for the initial condition of \Autoref{eq:Euler_eqs}, is possible with an MPS of bond dimension $1$.

Similarly to a discretized field, an operator $\hat{O}$ such as discretized differential operators 
$\frac{\partial}{\partial x}$ and $\frac{\partial}{\partial y}$ can also be decomposed into a product of tensors
($M^{i_{\kappa}^{j}.i_{\kappa}'^{j}}_{m_{j-1},m_j}$ with $\kappa \in \{ x,y \}$) like follows \cite{schollwoeckDensitymatrixRenormalizationGroup2011a}
\begin{align}
   \label{eq:QTToperator}
   \hat{O}  = \sum_{i_{x},i_{y}}
   \sum_{i_{x}',i_{y}'}
      M^{i_{x}^{(0)},i_{x}'^{(0)}}\ldots M^{i_{y}^{(n-1)},i_{y}'^{(n-1)}}
      \ket{(i_y',i_x')_b}
      \bra{(i_y,i_x)_b}.
\end{align}
Similar to the previous subsection, the summations over indices $m_j$ are omitted. We refer to $\hat{O}$ as a Matrix Product Operator (MPO).

For discretization of the differential operators over the coordinate, we adopt the finite difference formalism \cite{mitchell1980}.
The matrix elements for the first-order derivative in $x$ (second-order in accuracy) are given by
\begin{align}
   \hat{O}_{\partial x} 
   = \frac{1}{2} \sum_{i_x, i_y} \ket{(i_y,i_x)_b}\bra{(i_y,i_x + 1)_b} -\ket{(i_y,i_x)_b}\bra{(i_y,i_x - 1)_b} ,
\end{align}
where $\ket{(i_y, -1)_b} = \ket{(i_y, N-1)_b}$ and $\ket{(i_y, N)_b} = \ket{(i_y, 0)_b}$, and the operator in $y$ can be represented similarly.
Fortunately, one can immediately write tensors $M$ without constructing a full matrix and performing SVDs.
A low-rank MPO of $\hat{O}_{\partial x}$ ($\hat{O}_{\partial y}$) as well as $\hat{O}_{\partial xx}$ ($\hat{O}_{\partial yy}$) is well-known,
which we refer the readers to Refs.~\cite{kazeevLowRankExplicitQTT2012, kiffnerTensorNetworkReduced2023}. Note here that higher-order approximations are also possible with the bond dimension growing linearly with the order of the approximation ~\cite{garcia2021quantum}.

An MPO representation of the term responsible for the absorbing boundary condition, however, 
has not been discussed in the literature. 
We find that the sponge operator 
\begin{align}
   \hat{O}_{\text{sponge}} = \sum_{i_x,i_y} \gamma(\Delta i_x-L,\Delta i_y-L)\ket{(i_y,i_x)_b}\bra{(i_y,i_x)_b},
\end{align}
where $\gamma(x,y)$ is defined in \appref{app:sponge}, 
can be efficiently approximated with a low-rank MPO with the 
operator bond dimension $\chi_{\hat{O}}=4$ \cite{Siegl2024}.

\section{MPO-to-quantum circuit translation\label{app:MPOQC}}
To perform MPS-MPO contraction on a quantum computer, 
we also require a quantum mechanical representation of MPO. 
However, due to the unitarity of quantum mechanics, 
direct translation of general MPOs into quantum operators is not possible.
One must first, unitarize the operator by adding auxiliary qubits and probabilistically project out the unnecessary state to 
obtain the desired MPS-MPO contraction.

The translation of MPOs has been reported in \cite{Nibbi2024-arxiv, termanovaTensorQuantumProgramming2024}. 
The approach by Termonava et al.~will be of primary focus here, due to the favorable scaling of ancillary qubits with the bond dimension of the MPO, independent of the number of state-encoding qubits.

To compile an arbitrary MPO $\mathcal{M}$ with the maximum bond dimension $\zeta$ to quantum gates, 
we employ the strategy detailed in \cite{termanovaTensorQuantumProgramming2024}. 
We introduce an MPO  $\mathcal{Q}$, with site tensors $\mathcal{Q}[j]$, which should approximate $\mathcal{M}$ while obeying isometric constraints. 
Due to the restrictions imposed by these constraints, we need to expand the search for $\mathcal{Q}$ to a larger Hilbert space. To this end, we equip $\mathcal{Q}$ with a larger bond dimension than $\mathcal{M}$ by setting its bond dimension to 
$2^{z}>\zeta $  with an integer $z$.

To find optimal tensors of $\mathcal{Q}$, we solve the following optimization problem 
\begin{align}
   C(a_q,\mathcal{Q}) =\ &\underset{a_q,\mathcal{Q}}{\text{min}}
    \  \lVert a_q \mathcal{Q} -\mathcal{M} \rVert^2, \nonumber\\
       &\text{subject to} \ \ \mathcal{Q}[1]^{\dagger}  \in \text{St}(r, s) \notag\\  
       & \hspace{51pt}\mathcal{Q}[j] \hspace{1pt} \ \in \text{St}(r, s)  \quad \forall j = 2, \ldots, n, \notag
\end{align}
on a classical computer, under a condition that reshaped tensors of $\mathcal{Q}$ reside in Stiefel manifold \appref{app:MPOQC},
where $a_q$ is a normalization constant, $\lVert \cdot \rVert$ the Frobenius norm and 
$\text{St}(r, s)$ the Stiefel manifold \cite{Li2020}.
Here, we understand the tensors $\mathcal{Q}[j]$ as isometric matrices 
$\mathcal{Q}[j]:=\mathcal{Q}[j]^{(\alpha_{j-1}, \sigma_j), (\sigma_j', \alpha_{j})}$ upon reshaping, 
where $\alpha_{j-1}$ and $\alpha_j$ are bond indices of $Q[j]$. 
The normalization constant $a_q$ is computed with \cite{termanovaTensorQuantumProgramming2024}
\begin{align}
    a_q = \text{Re} \frac{\text{tr}\left[\mathcal{Q}^\dagger  \mathcal{M}  \right]}{\lVert \mathcal{M} \rVert^2}.
\end{align}
Finally, the full algorithm can be broken down into the following steps: (i) Initialize the isometric cores of $\mathcal{Q}$ 
(ii) Compute normalization $a_q$ 
(iii) Perform a single Riemannian gradient step on all tensors $Q[j]$ \cite{Li2020} 
(iv) Repeat (ii) and (iii) in an alternating manner until the error measure 
$\epsilon = \frac{||a_q\mathcal{Q}-\mathcal{M}||^2}{||\mathcal{M}||^2} $ reaches the set tolerance.
For a detailed explanation and discussion of this method, 
cf.~\cite{Siegl2024} and \cite{termanovaTensorQuantumProgramming2024}. %
As discussed in the main text, the tensors in $\mathcal{Q}$ and unitaries in $Q$ are related via a simple reshaping and projection 
$\mathcal{Q} = \ket{0}_{\mathrm{aux}}\bra{0}_{\mathrm{aux}} Q \ket{0}_{\mathrm{aux}}\bra{0}_{\mathrm{aux}} $. 
\section{Details of numerical simulations for problem case 1}\label{app:PC1numdetail}

For all the results with regards to case 1, we solve the time evolution of the pressure field under the wave equation derived from the 2-dimensional linearized Euler equation given in \Autoref{eq:discretized}.
We solve this equation in the realistic set of physical parameters for simulating aerodnyamics with ambient temperature and pressure, where density is $\rho=1.2250$ [kg/m\textsuperscript{3}] and the speed of sound is $c=340.2$ [m/s].

To reduce the computational resources, we utilize the 4-fold rotation symmetry of the model, 
where we only simulated one corner of the grid, including the center. 
The finite difference stencils are modified such that they implement the correct boundary conditions that respect this symmetry.

\subsection{Cost function}\label{app:costfunction}
The evolution of an unnormalized pressure field by one-time step, 
$\ket{p(x,y,t+\Delta t)}$ to 
$\ket{p(x,y,t+\Delta t)}=\lambda_p(t+\Delta t)U(\vec{\theta}_{\mathrm{opt}}(t+\Delta t))\ket{0}^{\otimes n}$, 
is achieved by optimizing a parameterized circuit $U(\vec{\theta})$ with a set of classical parameters 
$\theta_0$ and $\vec{\theta}=\{\theta_1,\theta_2, \ldots\}$.
With this formulation, $\ket{p(x,y,t+\Delta t)}$ is obtained by optimal $U(\vec{\theta})$ and $\theta_0$ 
that minimizes the cost function $\mathcal{C}$, 
constructed based on the discretization \Autoref{eq:discretized}, 
by replacing the discretized field and operators with quantum mechanical states and their normalization factors, 
and translated MPOs, their success probabilities, and their normalization factors
\begin{align}
   \label{eq:weq_costfun}
   \mathcal{C}(\theta_0,\vec{\theta}) =& 
   \Big|\Big|\theta_0 U(\vec{\theta})\ket{0} -\lambda_p(t)
   \mathcal{M}_1^{\kappa}
   \ket{\mathfrak{p}(x,y,t)} \nonumber\\
                                       &- \lambda_p(t-\Delta t) \mathcal{M}_2 \ket{\mathfrak{p}(x,y,t-\Delta t)} \nonumber\\
               &- 2c^2A_0\cos(2\pi f t) \sigma^x_{n_x}\sigma^x_{n_y}\ket{0}\Big|\Big|^2 \nonumber\\
               =& |\theta_0|^2 
               - 2\Re \Big\{ \theta_0\bra{0} U^{\dag}(\vec{\theta}) 
                  \Big( 
                     \lambda_p(t)q^{\kappa}_1
                     P_{\ket{0}_{\mathrm{aux}}\bra{0}_{\mathrm{aux}}}
                     Q_1^{\kappa} \ket{p(x,y,t)} \nonumber\\
                &- \lambda_p(t-\Delta t)q_2 P_{\ket{0}_{\mathrm{aux}}\bra{0}_{\mathrm{aux}}}
                Q_2 \ket{p(x,y,t-\Delta t)} \nonumber\\
               &- 2c^2A_0\cos(2\pi f t) \sigma^x_{n_x}\sigma^x_{n_y}\ket{0}\Big) \Big\} + \mathrm{Cst.},
\end{align}
where $\mathrm{Cst.}$ is an abbreviation of the terms that do not depend on $\theta_0$ or $\vec{\theta}$, 
$\mathcal{M}_1^{\kappa}$ and $\mathcal{M}_2$ are MPOs 
\begin{align}
   \mathcal{M}_1^{\kappa} &= 
   \left(2\mathbb{I} + (\Delta t)^2\left(c^2 O_{\plm{}{x}{2}}^{\kappa}  + c^2 O_{\plm{}{y}{2}}^{\kappa}  
   - \gamma(x,y) \right)\right) 
   = q_1^{\kappa}\mathcal{Q}_1^{\kappa},
   \nonumber\\
   \mathcal{M}_2 &= \left( \mathbb{I} - \Delta t \gamma (x,y)  \right)
         = q_2\mathcal{Q}_2,
\end{align}
with $q^{\kappa}_1$ and $q_2$ as the normalization constant of unitary approximated MPO $\mathcal{Q}^\kappa_1$ and $\mathcal{Q}_2$ 
respectively, 
$Q^{\kappa}_1$ and $Q^{\kappa}_2$ are their translated unitaries, and
$P_{\ket{0}_{\mathrm{aux}}\bra{0}_{\mathrm{aux}}}$ is the projection of the auxiliary qubits to $\ket{0}_{\mathrm{aux}}$ state,
and $\kappa$ is the order of finite difference approximations of the derivative operators. 
For translating the operators $Q^{k}_1$, we have the error measures of 
$\epsilon_{2}=4\times10^{-8}$, $\epsilon_{4}=5\times10^{-8}$, and $\epsilon_{6}=5\times10^{-8}$,
and for $Q_2$ we have $\epsilon = 3\times10^{-7}$. 

For the evaluation of the numerically exact evolution, $\ket{\mathfrak{p}(x,y,t)}_{\mathrm{exact}}$, 
we used the same operators $\mathcal{Q}_1^\kappa$ and  $\mathcal{Q}_2$ in the numerically exact form,
evaluated \ref{eq:discretized}, and obtained $\ket{\mathfrak{p}(x,y,t+\Delta t)}_{\mathrm{exact}}$. 

\subsection{Hadamard test}
\label{app:hadamardtest}
For the efficient evaluation of the expectation values in a cost function, we utilize the phase kickback algorithm 
\cite{aharonovPolynomialQuantumAlgorithm2009,lubaschVariationalQuantumAlgorithms2020}
by introducing one more ancillary qubit.
We evaluate the expectation value of a Hermitianized unitary operator $Q^\dag + Q$ 
by measuring the $\sigma^z$ expectation value of the ancilla qubit. Thus we only required to measure one qubit per evaluation of a term in a cost function.

The measurement of $\braket{Q^\dag +Q}$ via phase kickback on the ancilla is done by applying controlled $Q$
to a state $\ket{\Psi_0} \propto (\ket{0}_{\mathrm{anc}}+\ket{1}_{\mathrm{anc}})\ket{\psi}$ such that $Q$ is 
applied on $\ket{\psi}$ only if the state of the ancilla is $\ket{0}$. 
The resulting state is 
\begin{align}
   \ket{\Psi_1} =\frac{1}{\sqrt{2}}\ket{0}_{\mathrm{anc}}Q\ket{\psi} 
   + \frac{1}{\sqrt{2}}\ket{1}_{\mathrm{anc}}\ket{\psi}.
\end{align}
After applying the Hadamard gate $H$ on the ancilla, 
we obtain
\begin{align}   \braket{\sigma^z}_{\mathrm{anc}} = \bra{\Psi_1}H\sigma^{z}_{\mathrm{anc}} H \ket{\Psi_1} 
   =
   \Re \{ \bra{\psi}Q\ket{\psi} \}
   +
\frac{1}{2}\Re \{ \bra{\psi}(Q^\dag+ Q)\ket{\psi} \}.
\end{align} 

For evaluating the cost function $\mathcal{C}(\theta_0,\vec{\theta})$, 
we require the evaluation of the ancilla given a successful projection onto the state $\ket{0}_{\mathrm{aux}}$ 
of the auxiliary qubits.
Here we take the evaluation of the term 
$\bra{0}U^\dag(\theta(t+\Delta t))P_{\ket{0}_{\mathrm{aux}}\bra{0}_{\mathrm{aux}}}Q^{\kappa}_1U(\theta(t))\ket{0}$
as an example, 
where 
$\ket{\mathfrak{p}(x,y,t)}=U(\theta(t))\ket{0}$ and 
$\ket{\mathfrak{p}(x,y,t+\Delta t)}=U(\theta(t+\Delta t))\ket{0}$. 
This term is evaluated by first applying $H$ to the ancilla, 
then apply controlled $U(\theta(t))$ to the $n$ state qubits that encodes the field $p$, 
and then apply controlled $Q^{\kappa}_1$ to the $n$ qubits and $z$ auxiliary qubits with ancilla as a control qubit.
After these steps, we have a state 
\begin{align}
   \ket{\Psi} = \frac{1}{\sqrt{2}}\ket{0}_{\mathrm{anc}}Q^{\kappa}_1U(\theta(t))\ket{0} 
   + \frac{1}{\sqrt{2}}\ket{1}_{\mathrm{anc}}\ket{0},
\end{align}
where $\ket{0} = \ket{0}_{\mathrm{state}}\ket{0}_{\mathrm{aux}}$. 
Upon a successful projection on to the $\ket{0}_{\mathrm{aux}}$ state, the resulting state becomes 
\begin{align}
   \ket{\Psi'} = \frac{1}{\mathcal{N}} 
   \left(\frac{1}{\sqrt{2}}\ket{0}_{\mathrm{anc}}P_{\ket{0}_{\mathrm{aux}}\bra{0}_{\mathrm{aux}}}Q^\kappa_1U(\theta(t))\ket{0} 
   + \frac{1}{\sqrt{2}}\ket{1}_{\mathrm{anc}}\ket{0}\right),
\end{align}
where $\mathcal{N}=\sqrt{\frac{1}{2}P^{\kappa}_{1} + \frac{1}{2}}$ is a normalization factor with 
$P^{\kappa}_{1}$ being the success probability of a crude application of 
$Q^{\kappa}_1$ to the state $\ket{\frak{p}(x,y,t)}$ and obtaining $\ket{0}_{\mathrm{aux}}$, where we have
\begin{align}
   P^{\kappa}_{1} = \bra{\frak{p}(x,y,t)}(Q^{\kappa}_1)^\dag
   P_{\ket{0}_{\mathrm{aux}}\bra{0}_{\mathrm{aux}}}Q^{\kappa}_1\ket{\frak{p}(x,y,t)}. 
\end{align}
In the actual evaluation using the quantum computer, this projection is done by performing a projective measurement 
on auxiliary qubits. Upon measuring the auxiliary state $\ket{0}_{\mathrm{aux}}$, 
one proceeds to apply controlled $U^{\dag}(\theta(t+\Delta t))$ and then applies Hadamard gate to the ancilla. 
The resulting state is 
\begin{align}
   \ket{\Psi''}  = 
   \frac{1}{\mathcal{N}} \left(
   \frac{1}{\sqrt{2}}\ket{+}_{\mathrm{anc}}U^{\dag}(\theta(t+\Delta t))
   P_{\ket{0}_{\mathrm{aux}}\bra{0}_{\mathrm{aux}}}Q^\kappa_1U(\theta(t))\ket{0} 
   + \frac{1}{\sqrt{2}}\ket{-}_{\mathrm{anc}}\ket{0}\right),
\end{align}
where $\ket{\pm}$ are the $\pm1$ eigenstates of the $\sigma^{x}$ (Pauli-x) operator. 
The expectation value $\braket{\sigma^{z}_{\mathrm{anc}}}$ of this states is, therefore,
\begin{align}
   \braket{\sigma^{z}_{\mathrm{anc}}} &= 
   \frac{1}{\mathcal{N}^2}
   \frac{1}{2}\bra{\mathfrak{p}(x,y,t+\Delta t)}
   P_{\ket{0}_{\mathrm{aux}}\bra{0}_{\mathrm{aux}}}Q^{\kappa}_1\ket{\mathfrak{p}(x,y,t)}
   +
   \frac{1}{\mathcal{N}^2}\frac{1}{2}\bra{\mathfrak{p}(x,y,t)}
   (Q^{\kappa}_1)^\dag P_{\ket{0}_{\mathrm{aux}}\bra{0}_{\mathrm{aux}}}\ket{\mathfrak{p}(x,y,t+\Delta t)} \nonumber\\
                                      &= 
                                      \frac{1}{\mathcal{N}^2}\mathrm{Re} 
   \left\{\bra{\mathfrak{p}(x,y,t+\Delta t)}
   P_{\ket{0}_{\mathrm{aux}}\bra{0}_{\mathrm{aux}}}Q^{\kappa}_1\ket{\mathfrak{p}(x,y,t)} \right\},
\end{align}
as desired. 
Here, $\mathcal{N}^2$ is simply a probability of measuring $\ket{0}_{\mathrm{aux}}$,
and thus as we run this phase kickback circuit a number of times, the estimate of $\mathcal{N}$ improves alongside the
estimate of $\braket{\sigma^{z}_{\mathrm{anc}}}$. 

\subsection{Analytical result for $2D$ wave equation with a point source}\label{app:pc1analytical}
The wave equation \Autoref{eq:Euler_eqs_with_sponge}, where boundary is at infinity 
is analytically solvable, and is known to be analytically expressible in terms of 
Hankel functions and describes outgoing waves \cite{ostashevEquationsFinitedifferenceTimedomain2005}.
Using polar coordinates, in the limit $2\pi f r/c \gg 1$
we find an asymptotic expression for the amplitude of the pressure field fluctuation
\begin{equation} \label{asymptotic_solution}
   p(r, \theta) \simeq \frac{A_{0}}{2\pi}\sqrt{\frac{f}{cr}}.
\end{equation}
where $c$ is a speed of sound, $f$ is a source frequency, and $r$ is the radial distance from the source. 
In the other limit, the solution exhibits a logarithmic singularity 
($p \sim \mathcal{O}(\log r)$) for $r \rightarrow 0$ \cite{rienstra2023}.

\section{Solving Navier-Stokes equations on curvilinear coordinates with MPS and MPO}\label{app:NSCC}
\begin{figure}[t]
   \centering
   \includegraphics[scale=1.0]{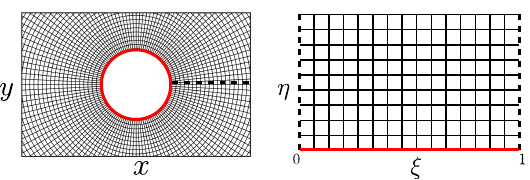}
   \caption{{\bf Depiction of curvilinear coordinates.} 
      We define our curvilinear coordinates parametrized by $\xi$ and $\eta$ in the physical domain (left)
      in such a way that the $\eta=0$ boundary of the orthogonal computational grid (right)
      corresponds to the boundary of an immersed object (red). 
      The direction $\xi$ is chosen such that its boundaries lie at the branch cut (dashed) with the periodic boundary condition. 
   \label{fig:compgrid}}
\end{figure}

In this section, we explain how time evolutions of the fields via the incompressible Navier-Stokes equations 
in 2D curvilinear coordinates are done to obtain our main results.
We first map points in the physical grid $(x,y)$ (\Autoref{fig:compgrid}, left) 
to an orthogonal computational grid $(\xi,\eta)$ (\figref{fig:compgrid}, right). 

\subsection{Chorin's projection}\label{app:Chorins} 
The incompressible Navier-Stokes equations are given by \cite{DEFRUTOS2016}
\begin{align}
    \nabla \cdot \bm{u} &= 0, \\
    \partial_t \bm{u}+(\bm{u} \cdot \nabla)\bm{u} &= -\frac{\nabla p}{\rho} + \nu \Delta \bm{u}, \label{eq:mom-eq2}
\end{align}
where $\bm{u}=(u,v)$ is the velocity field and $p$ the pressure. Here, $\rho$ refers to the density, $\nu$ to the kinematic viscosity.

To solve this initial value problem with initial velocity field vector $\bm{u}=\bm{u}_0$, Chorin's projection method can be employed. This method is explicit and belongs to the class of fractional-step methods \cite{Ferziger2002:CMFD}. At time step $k$, we denote our velocity field and pressure field by $\bm{u}^{(k)}$ and by $p^{(k)}$, respectively. To advance the flow fields by one-time step of size $\Delta t$, Chorin's method involves three main steps as follows
\begin{enumerate}
    \item[(1)] Computing the intermediate velocity field  $\bm{u}^*$ by solving the momentum equations, neglecting the pressure gradient term:
    \begin{equation}
        \bm{u}^* =  \bm{u}^{(k)} + \Delta t \cdot (-(\bm{u}^{(k)} \cdot \nabla)\bm{u}^{(k)} + \nu \Delta \bm{u}^{(k)}).
    \end{equation}
    \item[(2)] Solving the Poisson equation to determine $p^{(k+1)}$, where the source term (right-hand side of eq.) is given by the divergence of $\bm{u}^*$:
    
    \begin{equation}
       \Delta p^{(k+1)} = \frac{\rho}{\Delta t} \nabla \cdot \bm{u}^*. \label{eq:Poisson}
    \end{equation}

    \item[(3)] Divergence-free projection of the intermediate velocity field to arrive at the updated velocity field
       \label{item:temp3}
    \begin{equation} 
        \bm{u}^{(k+1)} = \bm{u}^* - \frac{\Delta t}{\rho} \nabla p^{(k+1)}.
    \end{equation}
\end{enumerate}

\subsection{Scaling}\label{app:Scaling}
Firstly, we have an offline step, the \textit{transformation step}, which needs to be performed once per system and grid, to create the \textit{metrics} and curvilinear differential operators. 
This step involves contractions of MPO-MPS in a pointwise manner. Hence, it scales as $\mathcal{O}(\chi_{met}^2)$, where $\chi_{met}$ is the maximum bond dimension of an MPS representing one of the \textit{metrics}. We note the bond dimension $\chi_{met}$ is grid-dependent. This scaling is straightforwardly seen, if the MPS describing the \textit{metric} is extended by a $\delta$-tensor as in \cite{gourianovQuantuminspiredApproachExploit2022}. Then it becomes a MPO-MPO pointwise multiplication, which can be done in a DMRG-like fashion \cite{schollwoeckDensitymatrixRenormalizationGroup2011a}, where the extended \textit{metric} MPO involves a bond dimension of $\chi_{met}$ and the Cartesian derivative $\partial_\xi$ of order unity.

Secondly, we have the online steps that must be performed in each Chorin's step. Here, the computationally most expensive operations in this fractional-step method are
\begin{enumerate}
    \item[I)] Computing the \textit{convective term} $(\bm{u} \cdot \nabla)\bm{u}$ in \Autoref{eq:mom-eq2} due to pointwise multiplication of two MPS, which scales as $ \underline{\mathcal{O}(\chi^3)}$ \cite{Michailidis2024}
\item[II)] Solving the \textit{Poisson equation} (Eq.~\ref{eq:Poisson}) for the pressure using a Density-Matrix-Renormalization-Group type approach scales as $\mathcal{O}(\chi^4)$, if local systems of equations of size $\chi^2 \times \chi^2$ are addressed by iterative methods, as the matrix-vector multiplication has a complexity of $\mathcal{O}(k^2)$ for $k \times k$ matrices. Oseledets et al.~\cite{Oseledets2012} have shown that it is not necessary to compose the local  $\chi^2 \times \chi^2$ matrices; instead, the matrix-vector product is computed directly by tensor contractions, which scale as $\underline{\mathcal{O}(\chi^3)}$.
\end{enumerate}
\end{document}